# Some Aspects of Rotational and Magnetic Energies for a Hierarchy of Celestial Objects

C Sivaram and Kenath Arun

Indian Institute of Astrophysics, Bangalore

**Abstract:** Celestial objects, from earth like planets to clusters of galaxies, possess angular momentum and magnetic fields. Here we compare the rotational and magnetic energies of a whole range of these celestial objects together with their gravitational self energies and find a number of interesting relationships. The celestial objects, due to their magnetic fields, also posses magnetic moments. The ratio of magnetic moments of these objects with the nuclear magnetic moments also exhibits interesting trends. We also compare their gyromagnetic ratio which appears to fall in a very narrow range for the entire hierarchy of objects. Here we try to understand the physical aspects implied by these observations and the origin of these properties in such a wide range of celestial objects, spanning some twenty orders in mass, magnetic field and other parameters.

Celestial objects over the entire range of mass and size scales possess intrinsic properties such as rotation (through their angular momentum) and magnetic fields. In case of many of these objects, such as galaxies, the origin of neither is well understood [1]. By comparing the rotational and magnetic energies of a whole range of these celestial objects together with their gravitational self energies we find a number of interesting relationships [2, 3].

The gravitational self energy associated with an object of mass M and radius R is given by:

$$E_{grav} = \frac{GM^2}{R} \qquad \ldots (1)$$

The magnetic field energy is given by:

$$E_{mag} = B^2 R^3 \qquad \ldots (2)$$

Where B is the average magnetic field.

The rotational energy is given by:

$$E_{rot} = \frac{1}{2} M \Omega^2 R^2 \qquad \ldots (3)$$

Where $\Omega = \frac{2\pi}{P}$; $P$ being the period of rotation

For example, in the case of earth, with $M = 6 \times 10^{27} g$, $R = 6 \times 10^8 cm$, $P \approx 24h$ and $B \sim 0.1 G$, we have the following:

$$E_{grav} = 4 \times 10^{39} \, ergs, \; E_{mag} = 2 \times 10^{24} \, ergs, \; E_{rot} = 6 \times 10^{36} \, ergs \qquad \ldots (4)$$

The ratio of the rotational and magnetic energies to the gravitational energy is therefore given as:

$$\frac{E_{mag}}{E_{grav}} \approx 3 \times 10^{-13} \text{ and } \frac{E_{rot}}{E_{grav}} \approx 10^{-3} \qquad \ldots (5)$$

These ratios for a hierarchy of celestial objects are tabulated below (for typical value of the parameter):

| Object | Rotation Period | Magnetic Field | $E_{mag}/E_{grav}$ | $E_{rot}/E_{grav}$ |
|---|---|---|---|---|
| Earth | 24 hours | 0.5 G | $\sim 10^{-13}$ | $10^{-3}$ |
| Jupiter | 10 hours | ~50 G | $10^{-14}$ | $10^{-1}$ |
| Sun | 25 Days | 10 G | $10^{-13}$ | $10^{-4}$ |
| White Dwarf | ~ 1 min | $\sim 10^6 G$ | $5 \times 10^{-13}$ | $3 \times 10^{-3}$ |
| Pulsar | ~ 1 s | $\sim 10^{12} G$ | $10^{-12}$ | $10^{-7}$ |
| ms Pulsar | $\sim 10^{-3} s$ | $\sim 10^9 G$ | $\sim 10^{-18}$ | $10^{-1}$ |
| Magnetar | ~ 1 s | $\sim 10^{15} G$ | $10^{-6}$ | $10^{-7}$ |
| Galaxy | 250 Myr | $\sim 10^{-6} G$ | $10^{-4}$ | $10^{-1}$ |
| Cluster | ~ 250 Myr | $\sim 10^{-6} G$ | $10^{-1}$ | $10^{-1}$ |

*Table 1: Ratios of rotational and magnetic energies to the gravitational energy*

Some comments are in order: We see that these ratios are comparable for magnetars and galaxies, even though the rotation and magnetic fields associated with pulsars are much higher than those associated with galaxies [4, 5].

For white dwarfs, with maximum magnetic fields seen $(\sim 10^6 - 10^7 G)$ and fast rotation periods (~1 minute), these ratios are again comparable. Even in the case of the sun and a typical pulsar the ratio of magnetic energy to gravitational energy is comparable [5]. The reason for this is that as the star collapses to form the compact pulsar, the magnetic flux is conserved, hence keeping the ratio constant. Thus $BR^2$ is a constant and $B^2 R^3$ goes as $1/R$, same as for gravitational self energy.

In the case magnetars the magnetic energy plays a bigger role. This probably indicates that the magnetic field of the star which collapsed was much larger to start off or the field was generated by some means (thermomagnetic, superconducting neutron fluid, etc.) [6].

In the case of galaxies, dark matter plays an important role as most of the galactic mass is due to this dark matter, which is not the case with pulsars. And dark matter does not couple with the magnetic field. So even though the fields are small, magnetic to gravitational field energy are not too different. Flat rotation curve (role of dark matter perhaps) ensures rotational energy is much closer to gravitational energy for both galaxies and clusters of galaxies than even in the case of rapid rotators like pulsars [3].

It has been noted that as far as rotational and gravitational energies are concerned, the scalings $J \sim M^2$ and $J \sim M^{5/3} \rho^{-1/6}$, play a universal role [2, 6, 7].

This may be understood in terms of stability of rotating bodies [8] which gives a limiting angular momentum:

$$J \geq 3 G^{1/2} M^{5/3} \rho^{-1/6} \qquad \ldots (6)$$

In a cluster of galaxies, the mass is dominated by dark matter. The mass goes as $M \propto R$. Therefore the gravitational energy, from equation (1), is given by $E_{grav} \propto R$

The magnetic energy is given by equation (2) as: $E_{mag} \propto R^3$

Therefore for a cluster we have:

$$E_{mag}/E_{grav} \propto R^2 \qquad \ldots (7)$$

The size of a galaxy cluster is ~30 times the size of a galaxy. This explains the ratio of magnetic energy to gravitational energy for a cluster being higher than that for a galaxy by a factor of $10^3$.

As mentioned earlier, in the case of galaxies, the origin of rotation and magnetic fields is not well understood. One possible scenario for the origin of rotation could be the transfer of angular momentum from the central black hole to the rest of the galaxy.

The horizon for a rotating (Kerr) black hole is given by:

$$r = m \pm \sqrt{m^2 - a^2} \qquad \ldots (8)$$

Here $m$ is the geometric mass and $a$ is the geometric angular momentum. From the condition that $r$ should be real, the limiting case is given by, $m = a$. That is:

$$\frac{GM}{c^2} = \frac{J_{MAX}}{Mc^2} \qquad \ldots (9)$$

From this, the maximum angular momentum is given by: [9]

$$J_{MAX} = \frac{M^2 G}{c} \qquad \ldots (10)$$

For a billion solar mass black hole, the maximum angular momentum is given by,

$$J_{MAX} \approx 10^{59} \, kgm^2/s \qquad \ldots (11)$$

In the case of a galaxy the rotational energy:

$$M\mathrm{v}^2 = \frac{GM^2}{R} \qquad \ldots (12)$$

The angular momentum of the galaxy is given by:

$$J_{gal} = M\mathrm{v}R = \frac{M\mathrm{v}^2 R}{\mathrm{v}} = \frac{GM^2}{\mathrm{v}} \qquad \ldots (13)$$

Where the galactic rotation velocity $\mathrm{v} \approx 250 \, km/s$.

In comparison for the black hole, the angular momentum is given by:

$$J_{BH} = \frac{M^2 G}{c} \qquad \ldots (14)$$

The mass of the galaxy, $M_{gal} \approx 10^3 M_{BH}$ and $v \approx 10^{-3} c$. This implies:

$$J_{gal} \sim 10^9 J_{BH} \qquad \ldots (15)$$

This excess angular momentum of the galaxy could be that transferred from the black hole to the ambient material, for before collapsing they would have to get rid of considerable angular momentum. For instance in the case of a 10 solar mass black hole, the maximum angular momentum allowed as per the above expression is of the order of:

$$J_{MAX} \approx 10^{43} kgm^2/s \qquad \ldots (16)$$

For a 10 solar mass star with 1.5 times the solar radius and an equatorial velocity of $v = 300 km/s$, its angular momentum is given by,

$$J = mvr \approx 10^{44} kgm^2/s \qquad \ldots (17)$$

We see that this angular momentum is larger than the maximum allowed angular momentum for a rotating black hole. Therefore, for these stars to end up as black holes, they will have to lose a considerable amount of their angular momentum. This angular momentum could be transferred to the host galaxy of the super massive black holes.

Although this alone is not sufficient to account for the total angular momentum of the galaxies. Evidently tidal interaction during close encounters of primordial galaxies forming in the early universe could account for their comparatively large angular momentum. Also dark matter haloes viriliazing and getting bound to seed structure could generate much angular momentum [10].

A way in which a massive star can lose the excess angular momentum before collapse into a black hole could be through gravitational quadrupole radiation. The gravitational power is given by:

$$P_{grav} = \frac{32}{5} \frac{G}{c^5} I_M^2 \Omega^6 \qquad \ldots (18)$$

As the star collapses to form the black hole the size ($R$) decreases as a function of time. Therefore both, the moment of inertia, given by $I_M = MR^2$, and the angular momentum, $J = I_M \Omega$, are dependent on time [11].

Also a black hole, according to 'no hair theorem', can have no magnetic field. One way in which it can lose the magnetic field before collapse is by dipole radiation, given by: [11]

$$P_{mag} = \frac{B^2 R^6 \Omega^4}{6c^3} \qquad \ldots (19)$$

As the star collapses, $R$ decreases with time. Hence flux conservation ($BR^2$ remains constant) implies that the magnetic field varies with time. Also $\Omega = \frac{v}{R}$ changes with time. Therefore:

$$\frac{dP_{mag}}{dt} = \frac{dP_{mag}}{dB} \frac{dB}{dt} \qquad \ldots (20)$$

Where we have:

$$\frac{dB}{dt} = \frac{dB}{dR} \frac{dR}{dt} \qquad \ldots (21)$$

The total power is then given by the integral of $\dot{P}_{mag}$ over the collapse time of the star. The celestial objects due to their magnetic fields posses' magnetic moments. The ratio of magnetic moments of these objects with the nuclear magnetic moments also exhibits interesting trends. The magnetic moment associated with the nucleus is given by:

$$\mu_N = \frac{\hbar e}{2m_N} \sim 10^{-23} \qquad \ldots (22)$$

The ratios of magnetic moments of the celestial objects $(BR^3)$ to the nuclear magnetic moments are compared to the mass ratio of these objects to the nucleon mass, i.e. the nucleon number.

| Object | Magnetic Moment $(BR^3)$ | $\mu/\mu_N$ | $M/M_N$ |
|---|---|---|---|
| Earth | $10^{26}\,Gcm^3$ | $2\times 10^{50}$ | $10^{51}$ |
| Jupiter | $10^{30}\,Gcm^3$ | $10^{54}$ | $10^{54}$ |
| Sun | $10^{34}\,Gcm^3$ | $10^{57}$ | $10^{57}$ |
| White Dwarf | $10^{34}\,Gcm^3$ | $10^{57}$ | $10^{57}$ |
| Neutron star | $10^{32}\,Gcm^3$ | $10^{55}$ | $10^{57}$ |
| Magnetar | $10^{33}\,Gcm^3$ | $10^{57}$ | $10^{57}$ |
| Galaxy | $10^{64}\,Gcm^3$ | $10^{87}$ | $10^{68}$ |
| Cluster | $10^{68}\,Gcm^3$ | $10^{91}$ | $10^{72}$ |
| Over Hubble Scale | $10^{75}\,Gcm^3$ | $10^{100}$ | $10^{78}$ |

*Table 2: Comparison of ratios of magnetic moment of celestial objects to the nuclear magnetic moment to mass ratio of these objects to the nucleon number*

So we see from the table that apart from galaxies and clusters of galaxies, for most other objects ranging from planets, giant planets and stars, including neutron stars and white dwarfs, the same ratio $\mu/\mu_N$ matches $M/M_N$, that is, ratio of their magnetic moment to the nucleon magnetic moment is the same as the number of baryons in these objects!

In the case of stars and magnetars there is a close correlation between the ratios of magnetic moments, $\mu/\mu_N$, and that of baryonic mass, $M/M_N$. But this is not true for galaxies and clusters and even the universe as a whole. This could be because in the case of these large scale structures it is not the baryonic mass but dark matter that dominates the total mass, which is not the case with stars. There is no dynamo in these cases which generates the magnetic field unlike that for stars.

In the case of these large scale structures, the ratio $\mu/\mu_N$ to $M/M_N$ seem to approach $e/\sqrt{G}m_e \approx 10^{21}$. The ratio $\mu/\mu_N$ for galaxy clusters to galaxies is the ratio of their masses!

The gyromagnetic ratio $(\gamma)$ of a particle or system is the ratio of its magnetic moment to its angular momentum. The gyromagnetic ratios for a hierarchy of celestial objects are tabulated below.

| Object | Magnetic Moment $(BR^3)$ | Angular Momentum $(MR^2\Omega)$ | $\gamma$ |
|---|---|---|---|
| Earth | $10^{26}\, Gcm^3$ | $2\times 10^{41}\, gcm^2 s^{-1}$ | $10^{-15}$ |
| Jupiter | $3\times 10^{30}\, Gcm^3$ | $2\times 10^{46}\, gcm^2 s^{-1}$ | $10^{-16}$ |
| Sun | $10^{34}\, Gcm^3$ | $6\times 10^{49}\, gcm^2 s^{-1}$ | $2\times 10^{-16}$ |
| White Dwarf | $10^{34}\, Gcm^3$ | $10^{50}\, gcm^2 s^{-1}$ | $10^{-16}$ |
| Pulsar | $10^{30}\, Gcm^3$ | $10^{46}\, gcm^2 s^{-1}$ | $10^{-16}$ |
| ms Pulsar | $10^{27}\, Gcm^3$ | $2\times 10^{49}\, gcm^2 s^{-1}$ | $5\times 10^{-23}$ |
| Galaxy | $10^{64}\, Gcm^3$ | $10^{79}\, gcm^2 s^{-1}$ | $10^{-15}$ |
| Cluster | $10^{68}\, Gcm^3$ | $2.5\times 10^{82}\, gcm^2 s^{-1}$ | $4\times 10^{-15}$ |

*Table 3: Gyromagnetic ratios for a hierarchy of celestial objects*

From the table we see that over the whole hierarchy of celestial objects the gyromagnetic ratio is almost remarkably constant except in couple of cases [2, 3, 7]. In the case of millisecond pulsar the magnetic fields are much weaker (recycled pulsar) and angular momentum is higher due to its faster spin. For Jupiter, as it is spinning close to the breaking limit, its angular momentum is greater. Hence its gyromagnetic ratio is an order less than that of the other celestial objects.

The constancy of the gyromagnetic ratio could be attributed to the fact that magnetic moment given by, $\mu = \dfrac{\sqrt{G}}{c} J$, gives the gyromagnetic ratio as: [2, 3]

$$\gamma = \frac{\mu}{J} = \frac{\sqrt{G}}{c} \approx 10^{-15} \qquad \ldots (23)$$

This roughly corresponds to the gyromagnetic ratios of the celestial bodies.

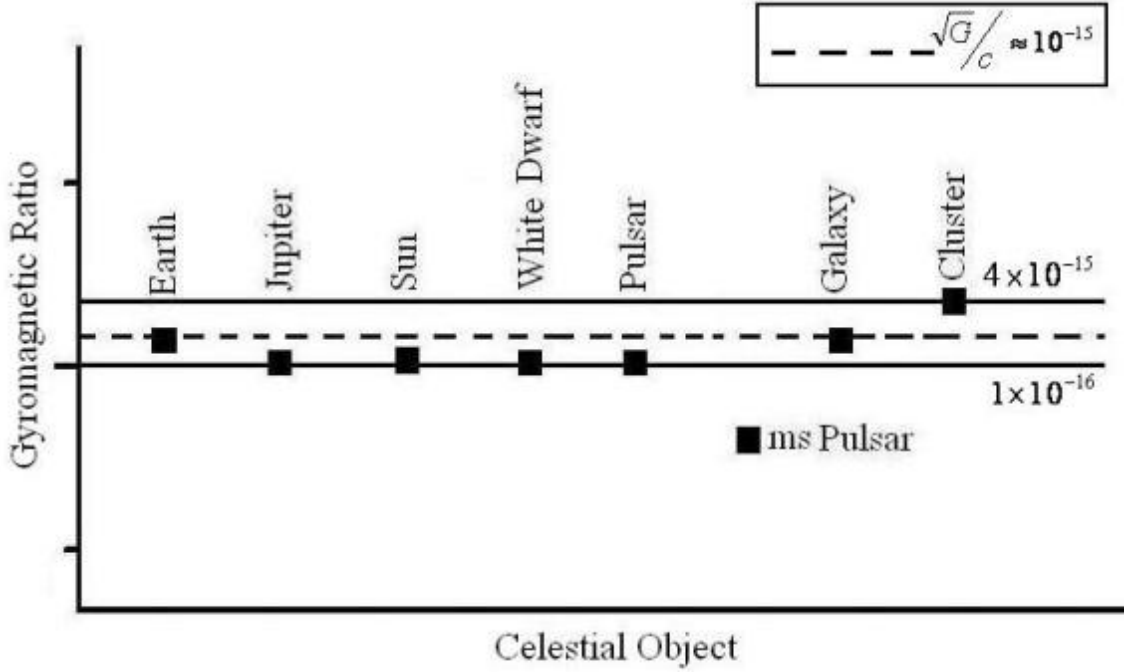

*Figure 1: Gyromagnetic ratio for a hierarchy of celestial objects over a range of 20 orders of magnitude of mass and magnetic field*

In the case of Mercury, even though it has a slow rotation (with a period of ~ 60 days) it is found to have a magnetic field. The above discussion could shed light as to why Mercury possesses this magnetic field. The angular momentum for mercury is given by:

$$J = MR^2\Omega \sim 2\times 10^{37}\ gcm^2 s^{-1} \qquad \ldots (24)$$

From equation (23), the corresponding magnetic field is given by:

$$B = \frac{\mu}{R^3} = \frac{\sqrt{G}}{c}\frac{J}{R^3} \approx 3\times 10^{-4}\ G \qquad \ldots (25)$$

This agrees with the observations from MESSENGER [12].

In the case of the electron, the Bohr magneton is given by:

$$\mu_B = \frac{\hbar e}{2m_e c} \qquad \ldots (26)$$

The ratio of electron magnetic moment to its angular momentum is thus given by:

$$\frac{\mu_B}{\hbar} = \frac{e}{2m_e c} \sim 10^7 \qquad \ldots (27)$$

The ratio of gyromagnetic ratio of the electron to that of celestial bodies is:

$$\frac{\gamma_e}{\gamma_{cel}} = \frac{e/2m_e c}{\sqrt{G}/c} = \frac{e}{\sqrt{G}m_e} \approx 10^{21} \qquad \ldots (28)$$

($\sqrt{G}m_e$ is the 'gravitational charge' of the electron)

So in a sense, as gravity and electromagnetism are the two long range forces, it is perhaps natural that the gyromagnetic ratios for astrophysical bodies and for atomic systems involve just $e/\sqrt{G}m_e$.

In general we can write for the magnetic moment:

$$\mu = \frac{\hbar e}{2m_P c} \frac{J}{\hbar} \qquad \ldots (29)$$

The angular momentum depends on the mass (equation (10)) given by:

$$J = pM^2 \qquad \ldots (30)$$

Where $p \approx 10^{-15} g^{-1} cm^2 s^{-1}$

And from equation (24) the magnetic moment is related to the angular momentum as:

$$J = q\mu \qquad \ldots (31)$$

Where $q \approx 10^{15} g^{1/2} cm^{-1/2}$

From this we can construct two dimensionless constants with two sets of constants $(G, c, p)$ and $(G, c, q)$. That is: [3]

$$\beta' = \frac{G}{cp} \sim 3 \times 10^{-3} \qquad \ldots (32)$$

$$\gamma' = \frac{c^2}{q^2 G} \sim 7 \times 10^{-3} \qquad \ldots (33)$$

It is interesting that these match the value of $\alpha \left( = \frac{e^2}{\hbar c} \right)$, the electromagnetic fine structure constant [3]. Thus:

$$p \sim \frac{G}{\alpha c} \approx 10^{-15} g^{-1} cm^2 s^{-1} \qquad \ldots (34)$$

$$q \sim \frac{c}{(\alpha G)^{1/2}} \approx 10^{15} \, g^{1/2} cm^{-1/2} \qquad \ldots (35)$$

We can relate these two quantities through the following:

$$\frac{q^2}{p} = \frac{c^3}{G^2} = \left(\frac{c^2}{G}\right)^2 \frac{1}{c} \qquad \ldots (36)$$

Again it may be of some interest to interpret $c^2/G$ as the superstring tension $\left(\sim \beta c^2/G, \beta \sim 1\right)$ [7, 13].

In a recent paper [14] there was a report on an extension of a scaling law relating the energy flux and the magnetic field strength for celestial objects ranging from planets to low mass stars. It is observed therein that the magnetic fields of stars follow the relation:

$$B \propto q_0^{2/3} \rho^{1/3} \qquad \ldots (37)$$

Where, $q_0$ is the flux and $\rho$ the density.

Here we would like to point out that the same type of scaling also holds true for galaxies. In the case of galaxies the density $\rho_{gal} \approx 10^{-24} \rho_{star}$ and the luminosity of a galaxy is typically $L_{gal} \approx 10^{12} L_{star}$. This scaling (equation (37)) implies $B_{gal} \sim 10^{-6} G$, as observed!

**Reference:**

1. V. Trimble, Comments Astrophysics, 10, 27, 1984
2. C. Sivaram, in Relativistic Astrophysics and Cosmology, Edited by V. de Sabbata and T. Karade, World Scientific, Singapore, 1984, p.228
3. V. de Sabbata and C. Sivaram, Nuovo Cimento, 100A, 919, 1988
4. P. Brosche, in Cosmology and Gravitation, Spin, Torsion, Rotation, Edited by P. E. Bergmann and V. de Sabbata, Plenum Press, N.Y, 1980
5. V. de Sabbata and M. Gasperini, Lett. Nuovo Cimento, 38, 93, 1983
6. P. S. Wesson, Astronomy and Astrophysics, 80, 296, 1979

7. C. Sivaram, Nature, <u>327</u>, 508, 1987
8. Landau and Lifshitz, Classical Theory of Fields
9. C. Misner, K. Thorne and J. Wheeler, Gravitation, Freeman and Co., 1973
10. Binzey and Tremaine, Galactic Dynamics
11. White Dwarf, Neutron Star and Black Hole, Shapiro and Tuekolsky
12. M. E. Purucker et al, 40th Lunar and Planetary Science Conference, (Lunar and Planetary Science XL), 2009, Texas, id.1277
13. L. J. Tassie, Nature, <u>323</u>, 40, 1986
14. U. R. Christensen, V. Holzwarth and A. Reiners, Nature, <u>457</u>, 167, 2009